\documentclass[11pt]{article}

\usepackage[utf8]{inputenc}
\usepackage[T1]{fontenc}
\usepackage{graphicx}
\usepackage{multirow}
\usepackage{amsmath,amssymb,amsfonts}
\usepackage{amsthm}
\usepackage{mathrsfs}
\usepackage[title]{appendix}
\usepackage{xcolor}
\usepackage{textcomp}
\usepackage{booktabs}
\usepackage{bm}
\usepackage{algorithm}
\usepackage{algorithmicx}
\usepackage{algpseudocode}
\usepackage{listings}

\usepackage[a4paper, total={6.5in, 9in}]{geometry}
\usepackage[numbers]{natbib}
\usepackage{url} 
\usepackage{hyperref}

\theoremstyle{plain}

\theoremstyle{definition}

\theoremstyle{remark}

\title{Laboratory-based x-ray microtomography with directional dark-field sensitivity}

\author{
\textbf{Carlos Navarrete-Le\'{o}n}$^{1,2}$\thanks{carlos.leon.17@ucl.ac.uk}, 
\'{A}lvaro Jos\'{e} Gonz\'{a}lez-Grajales$^{1,2}$, \\
Harry Allan$^{1,2}$, 
Adam Doherty$^{1,2}$, 
Alissa Parmenter$^{1,3}$, \\
Rocco D'Antuono$^{4,5}$, 
David Bate$^{1,6}$, 
Alberto Astolfo$^{1}$, \\
Silvia Cipiccia$^{1}$, 
Charlotte K. Hagen$^{1}$, 
Alessandro Olivo$^{1}$, 
\textbf{Marco Endrizzi}$^{1,2}$\thanks{m.endrizzi@ucl.ac.uk} \\
\\
\small $^{1}$Department of Medical Physics and Biomedical Engineering, University College London, \\
\small London WC1E 6BT, UK \\
\small $^{2}$X-ray Microscopy and Tomography Lab, The Francis Crick Institute, London NW1 1AT, UK \\
\small $^{3}$Department of Mechanical Engineering, University College London, London WC1E 6BT, UK \\
\small $^{4}$Crick Advanced Light Microscopy STP, The Francis Crick Institute, London NW1 1AT, UK \\
\small $^{5}$Department of Biomedical Engineering, University of Reading, Reading RG6 6AY, UK \\
\small $^{6}$Nikon X-tek Systems Ltd, Tring HP23 4JX, UK
}

\date{} % Leave empty or add date

\begin{document}
\maketitle

\abstract{We demonstrate dark-field x-ray microtomography in a compact, laboratory-based system capable of resolving attenuation, phase, and anisotropic scattering signals with micrometer-scale resolution across centimetre-scale samples. The method is based on two-directional beam tracking (2DBT), which requires only a single optical element and is compatible with standard x-ray sources and detectors. We validate the system’s capabilities through imaging of a custom-built phantom, a fibre-reinforced composite and ex-vivo biological tissues, including a bovine intervertebral disc, a rat heart, and a porcine meniscus. The results show that dark-field tomography provides complementary information to attenuation as well as to phase tomography, by revealing sub-resolution features such as fibre orientation and microstructural heterogeneity at length scales that are well below the voxel size. A key element of our system is its sensitivity to scattering along two orthogonal directions in the image plane, enabling the measurement of scattering anisotropy with a single exposure. As well as simple and robust, our approach is sensitive and precise. These findings demonstrate the potential of 2DBT for non-destructive and three-dimensional structural characterisation of samples and materials in engineering, materials science and biomedical applications.}

\section{Introduction}

X-ray microtomography is a powerful technique for non-destructive three-dimensional imaging of samples, with applications across biomedical research, materials science, and industrial metrology \cite{Withers2021}. Its achievable spatial resolution is typically constrained by the choice of system components, namely source emission spot size, detector pixel size and scintillator, and system magnification \cite{Carmignato2018}. Depending on the configuration, current systems can achieve voxel sizes ranging from several tens of nanometres to the micrometre scale. However, due to limitations in the number of detector pixels in the matrix and data throughput, there is often a trade-off between spatial resolution and the achievable field of view (FOV). For instance, pixel sizes on the order of 1~\textmu m typically restrict imaging to millimetre-scale samples, while centimetre-scale samples are limited to resolutions in the tens of microns.

X-ray dark-field (XDF) imaging can address this limitation by detecting small-angle scattering from unresolved microstructures within the sample \cite{Davis1995, Wilkins1998, Rigon2007, Pfeiffer2008}. By quantifying these scattering effects, XDF provides indirect access to sub-resolution structural information across a much larger FOV. For example, the orientation and spatial distribution of micrometre-scale fibres can be characterised within centimetre-sized objects, without the need to resolve individual fibres and subsequently trace their direction in space \cite{Jensen2010}. Dark-field imaging has been used to investigate sub-resolution processes in a variety of systems, including the formation of ice crystals in soft tissues during freezing \cite{John2024}, hydration fronts in cement paste \cite{Prade2016}, microfractures in fiber-reinforced composites \cite{Shoukroun2020}, the organization of axonal fibers in the brain \cite{Wieczorek2018}, and alveolar damage in irradiated lung tissue \cite{Burkhardt2021}.

Although laboratory-based XDF methods have been demonstrated, namely Talbot–Lau interferometry \cite{Pfeiffer2006, Pfeiffer2008}, edge illumination \cite{Endrizzi2014, Olivo2021}, speckle-based imaging \cite{Zanette2014}, random intensity modulation \cite{Magnin2023}, dual phase \cite{Kagias2017} and circular grating interferometry \cite{kagias2019}, their use has so far remained restricted to a limited number of research groups. Compared to synchrotron implementations, laboratory XDF systems have not yet seen widespread adoption. We think that this is largely due to challenges related to the requirements on the spatial coherence, on the detector resolution, or on the need for complex optical arrangements. Although each of the approaches listed above offers its own distinct set of advantages, we aim at further improving the uptake and application of XDF imaging methods.

In this work, we focus on two-directional \cite{ Dreier2020_2, NavarreteLeon2023} beam tracking (2DBT) \cite{Vittoria2015}, an approach that uses a single intensity modulator to form an array of independent beamlets. By analysing changes in beamlet position and shape, we retrieve, within a single exposure, attenuation, phase, and directional dark-field signals. 2DBT provides directional sensitivity without the need for multiple optical elements or repeated exposures, while also relaxing source and detector constraints on the system's spatial resolution \cite{Diemoz2014, Zekavat2024, NavarreteLeon2023_2}. The minimal hardware requirements (one modulator with a standard x-ray source and detector) allow implementation in a compact laboratory system with a footprint of less than one metre.

Here we report on laboratory-based dark-field radiography and micro-tomography using 2DBT. We demonstrate that this compact approach enables volumetric characterisation of fibrous and heterogeneous samples across centimetre-scale fields of view. We apply the methodology to a custom built phantom, a carbon fibre composite, a wood sample, and ex-vivo biological tissues, including a bovine intervertebral disc, a rat heart, and a porcine meniscus. Our results demonstrate that 2DBT can resolve fibre orientation and microstructural heterogeneity across a variety of materials, both hard and soft, and that this information remains inaccessible to attenuation or phase contrast alone. As well as illustrating this approach on an engineering phantom that is simple to interpret, we extend XDF imaging and anisotropic scattering analysis to soft tissues for applications in the biomedical sciences.

\begin{figure*}
\centering
\includegraphics[width=\linewidth]{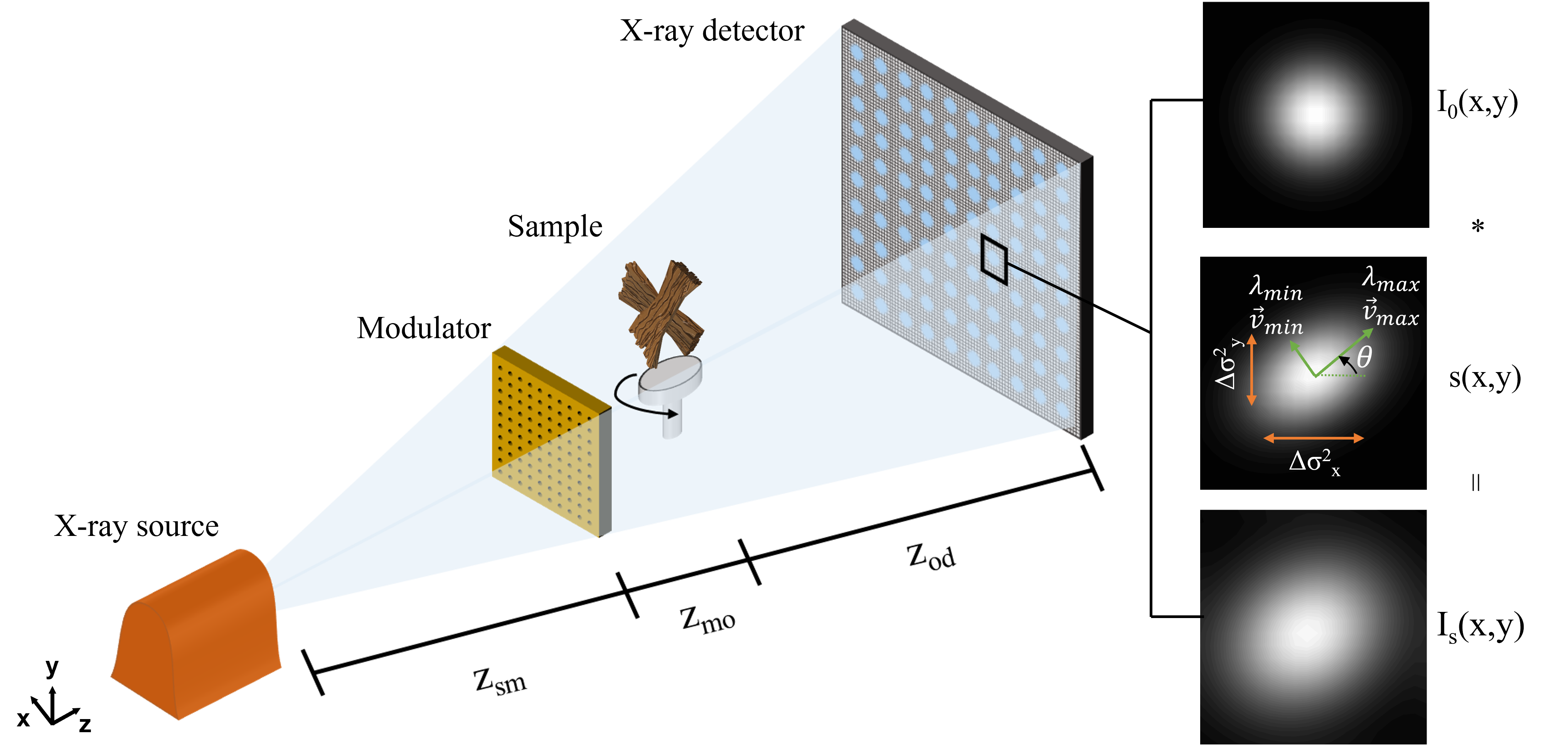}
\caption{\textbf{Schematic of the laboratory set-up for dark field imaging}. The system features a sealed microfocus x-ray tube, an intensity modulator, a sample stage, and an x-ray detector. The local scattering of the sample is described using a convolution model, where the intensity distribution of each probe, $I_s(x, y)$, is a result of the scattering function $s$ convolved with the probe distribution in the absence of the sample $I_0$. The two-dimensional sensitivity of the system allows retrieval of the variances ($\Delta\sigma^2_{x}$, $\Delta\sigma^2_{y}$) and the orientation of the scattering function $s(x,y)$. To illustrate the scattering effect, the probes shown here were modelled with high oversampling; in experimental conditions, each probe is sampled with 5 to 10 pixels, depending on the modulator used.}
\label{fig:system_darkfield}
\end{figure*}

\section{Materials and methods}

\subsection{Lab-based dark-field imaging setup}

The imaging system consists of a microfocus x-ray source, a modulator, and two types of detectors. The source is a sealed tungsten-target tube (L12161-07, Hamamatsu), operated at 40 kVp and 10 W. The modulator is fabricated from a 100~\textmu m-thick tungsten foil (Goodfellow) using laser ablation, which was used to form a regular array of circular apertures. Each aperture has a conical profile, with diameters of approximately 15~\textmu m on the source-facing side and 30~\textmu m on the detector-facing side, as measured using an optical microscope. Two different pitches were tested: 50~\textmu m and 100~\textmu m.

Images were acquired using two detector technologies. A photon-counting detector (MerlinX, Quantum Detectors) based on a Medipix3RX chip was employed for image quality benchmarking; it features a 55~\textmu m pixel pitch and a $1.4 \times 1.4~$cm$^2$ active area. A flat-panel detector (Hamamatsu C9732DK-11) with 50~\textmu m pixel pitch and a $12 \times 12$~cm$^2$ active area was used for large field-of-view radiography and tomography. Whilst the photon-counter offers quantum-limited noise performance it remains constrained to a small imaging area, on the other hand, the CMOS-based flat panel, covers much larger areas albeit with a trade-off in noise performance.

A schematic of the imaging system is shown in Figure~\ref{fig:system_darkfield}, which illustrates the arrangement of the x-ray source, modulator, sample stage, and detector. The standard imaging geometry used a source-to-modulator distance of $z_{sm} = 140$~mm, a modulator-to-object distance of $z_{mo} = 20$~mm, and an object-to-detector distance of $z_{od} = 540$~mm. For context, previous single-optical-element, direction-sensitive laboratory XDF setups reported bench lengths ranging 0.43–3.6 m \cite{Zanette2014, Magnin2023, Smith2023, Slyamov2022}; our implementation operates at $<$1 m and supports multi-contrast tomography.

\subsection{Model and signal retrieval}

The modulator structures the x-ray cone-beam into an array of spatially separated probes (beamlets), which traverse the sample and are subsequently resolved by the detector pixels. The local interaction of each beamlet with the sample can be described by a convolution model, where the recorded intensity distribution $I_s(x, y)$ is obtained by convolving the unperturbed intensity distribution  $I_0$ with a scattering function $s$, along with a shift due to refraction and an overall attenuation \cite{Endrizzi2014}:
\begin{equation}
I_s(x, y) = t (I_0 * s)(x - \Delta x, y - \Delta y),
\end{equation}
where $t$ is the local transmission, and $\Delta x$, $\Delta y$ are the lateral displacements of the beamlet on the detector plane induced by sample refraction. The scattering function $s(x, y)$ represents the two-dimensional probability distribution for an x-ray to be scattered from its original trajectory into point $(x,y)$. In this work, we approximate $s(x, y)$ with a Gaussian function. For samples exhibiting more complex scattering distributions, a multi-Gaussian model can be used to preserve the convolutional form of the model while capturing a wider range of characterstic scattering functions \cite{Endrizzi2014}.

The attenuation, refraction, and scattering signals are retrieved by comparing the probe intensity distributions with ($I_s$) and without ($I_0$) the sample. Transmission is obtained from the ratio of integrated intensities: $t = \frac{\sum_{i,k} I_s(x_i, y_k)}{\sum_{i,k} I_0(x_i, y_k)}$, where $x_i$ and $y_k$ index the detector pixels within a single beamlet window. Refraction is measured from the beamlet centroid displacements using a subpixel registration algorithm \cite{NavarreteLeon2023, Guizar2008}.
Scattering is quantified by measuring the increase in spatial variance of the beamlet intensity profile. For instance, the variance along $x$ is given by:

\begin{equation}
\sigma_x^2 = \frac{\sum_{i,k} (x_{i} - \bar{x})^2 \cdot I(x_{i},y_{k})}{\sum_{i,k}I(x_{i},y_{k})},
\end{equation}
where $\bar{x}$ is the centroid of the beamlet in the $x$-direction. The dark-field signal is then computed as:
$\Delta \sigma_x^2= \sigma_{x, s}^2 - \sigma_{x, 0}^2$. Analogous expressions hold for $\sigma_y$.

The two-dimensional sensitivity of the system enables directional scattering analysis through the covariance matrix of each beamlet \cite{Dreier2020}:
\begin{equation}
\bm{\Sigma} =
\begin{pmatrix}
\Delta \sigma_x^2 & \Delta \sigma_{xy} \\
\Delta \sigma_{xy} & \Delta \sigma_y^2
\end{pmatrix},
\end{equation}
where $\Delta \sigma_x^2$ and $\Delta \sigma_y^2$ are the increases in variance along the $x$ and $y$ directions, respectively, and $\Delta \sigma_{xy}$ is the change in covariance relative to the unperturbed intensity distribution computed as:

\begin{equation}
    \sigma_{xy} = \frac{\sum_{i,k} (x_i - \bar{x}) (y_k - \bar{y}) \cdot I(x_i,y_k)}{\sum_{i,k} I(x_i,y_k)}.
\end{equation}

The eigenvalues and eigenvectors of $\bm{\Sigma}$, obtained by eigendecomposition, correspond to the scattering strengths along the principal axes and their associated orientations (see beamlets in Figure \ref{fig:system_darkfield}). Let $\bm{v}_{\text{max}} = [v_x, v_y]$ be the eigenvector associated with the largest eigenvalue $\lambda_{\text{max}}$. The dominant scattering direction $\theta$ is then given by:
\begin{equation}\label{eq:theta}
    \theta = \arctan\left(\frac{v_y}{v_x}\right).
\end{equation}
The degree of anisotropy is quantified as:
\begin{equation}\label{eq:kappa}
    \kappa = \sqrt{1 - \left(\frac{\lambda_{\text{min}}}{\lambda_{\text{max}}}\right)^2},
\end{equation}
where $\lambda_{\text{min}}$ is the smallest eigenvalue. This coefficient ranges from 0 (isotropic scattering) to 1 (anisotropic scattering). Directional scattering analysis is performed independently for each beamlet, which enables pixel-wise mapping of scattering magnitude, direction, and anisotropy across the sample.

The retrieved multi-contrast signals are suitable for tomographic reconstruction. Transmission and refraction are treated as line integrals of the linear attenuation coefficient $\mu$ and the refractive index decrement $\delta$, respectively \cite{NavarreteLeon2023}:

\begin{equation}
\label{eq:mu}
-\ln t(x,y) = \int_L \mu(x', y', z'), dz,
\end{equation}

\begin{equation}
\label{eq:delta}
-\frac{\Delta \Phi(x,y)}{k} = \int_L \delta(x', y', z'), dz,
\end{equation}

where $\Delta \Phi(x,y)$ is the retrieved phase shift and $k$ is the wave-number of the x-ray field. The phase shift $\Delta \Phi(x,y)$ is obtained by integrating the phase gradients along $x$ and $y$, which are proportional to the beamlet displacements $\Delta x$ and $\Delta y$, respectively \cite{NavarreteLeon2023}. The line integrals are computed along the x-ray beam path $L$, which corresponds to the straight-line trajectory from the source source to each detector pixels, for every viewing angle.

Similarly, dark-field signals $\Delta \sigma_x^2$ and $\Delta \sigma_y^2$ are treated as line integrals of the directional scattering coefficients $\epsilon_x$ and $\epsilon_y$, which describe the angular broadening of the scattering function per unit length along the $x$ and $y$ directions \cite{Endrizzi2017, Doherty2023}:

\begin{equation}
\label{eq:sigma_tomo}
\Delta \sigma_x^2(x,y) = \int_L \epsilon_x(x', y', z'), dz, \quad
\Delta \sigma_y^2(x,y) = \int_L \epsilon_y(x', y', z'), dz.
\end{equation}

These relationships enable the reconstruction of volumetric maps of attenuation, phase, and scattering using standard tomographic algorithms applied to projection data acquired over multiple viewing angles.

\subsection{Experiments}

The system was first validated using a custom phantom composed of a carbon fibre bundle loop, a section of bamboo wood, and a paper sample. Carbon fibres and bamboo were selected as representative examples of anisotropic scatterers, as their internal microstructure is highly ordered. In ordered systems the X-ray scattering is expected predominantly in the direction orthogonal to the main orientation of the sub-resolution features where scattering originates. In contrast, the paper was selected as a control sample, representing isotropic scattering. Due to its composition of randomly oriented microfibres, the X-rays are scattered in all directions with the same magnitude. For such materials, scattering is expected to be uniform regardless of sample orientation. The phantom was imaged using both the photon-counting and flat-panel detectors in combination with the 100~\textmu m-pitch modulator. To improve sampling, the sample was scanned over a $8 \times 8$ grid with 15~\textmu m steps, which corresponds roughly to the modulator period projected onto the sample. At each grid position, 32 frames were acquired with 1~s exposure time per frame. Flat-field and dark-current images were also acquired, and the frames were averaged to increase signal-to-noise ratio.

For tomography, we first imaged a fibre-reinforced polymer composite, chosen for its relevance in non-destructive testing and its low atomic number composition. Low-atomic number elements and materials are associated with weak absorption contrast in the X-ray regime, and thus represent a challenge for conventional microtomography techniques. The scan was performed using the flat-panel detector and the 50~\textmu m-pitch modulator. A total of 2000 projections were acquired over 360$^\circ$ at each modulator position. To increase spatial sampling, the modulator was stepped by  12.5~\textmu m, for a total of $4 \times 4$ scans with 1.2~s exposure per frame, resulting in a total exposure time of 2000 $\times$ 4 $\times$ 4 $\times$ 1.2~s = 10.7~h.

To further assess the applicability of the system to biological and medical specimens, we imaged three ex-vivo soft tissues samples, where the microstructure of the fibres supporting the microanatomy are too small to be directly resolved by the imaging system, yet the dark-field contrast channel can reveal their presence and quantify their orientation. The samples were imaged using the flat-panel detector and the 100~\textmu m-pitch modulator. In radiography mode, a bovine intervertebral disc dissected from an ox tail was scanned. The modulator was stepped over an 8 $\times$ 8 grid of 12.5~\textmu m steps, and 2~s frames were acquired per step. For tomographic imaging, we imaged a porcine medial meniscus and a rat heart. For the meniscus, 1600 projections were acquired over 360$^\circ$. The modulator was stepped by  16.7~\textmu m steps in $x$ and $y$, for a total of 6 $\times$ 6 scans with 1.2~s exposure per frame, resulting in a total scan time of 1600 $\times$ 6 $\times$ 6 $\times$ 1.2~s = 19.2~h. For the rat heart, 1400 projections were acquired under identical conditions, with the modulator stepped by 20~\textmu m over a 5 $\times$ 5 grid, resulting in a total exposure time of approximately 1400 $\times$ 5 $\times$ 5 $\times$ 1.2~s = 11.7~h.

Sample preparation for the intervertebral disc and meniscus followed a modified dehydration protocol for cartilaginous tissue \cite{Kestil2018}. After dissection, samples were rinsed in 1\% phosphate-buffered saline (PBS) for 1h, fixed overnight in 70\% ethanol, and dehydrated through a graded ethanol series (70\textendash100\%). Samples were then immersed in hexamethyldisilazane (HMDS) for 24h and air-dried in a laminar flow hood. Dried samples were stored in airtight containers until imaging to prevent the sample from moisturizing. The rat heart was obtained from a 300~g Sprague–Dawley rat, euthanized under Schedule 1 protocols. It was fixed in paraformaldehyde (PFA) and prepared by critical point drying, as described in \cite{Savvidis2022}.

All reconstructions were computed using the Feldkamp–Davis–Kress algorithm implementation in CUDA of the Astra Toolbox \cite{Astra2015}. The volume renderings presented were obtained using the Avizo software (Thermo Fischer Scientific).

\begin{figure*}[h]
\centering
\includegraphics[width=\linewidth]{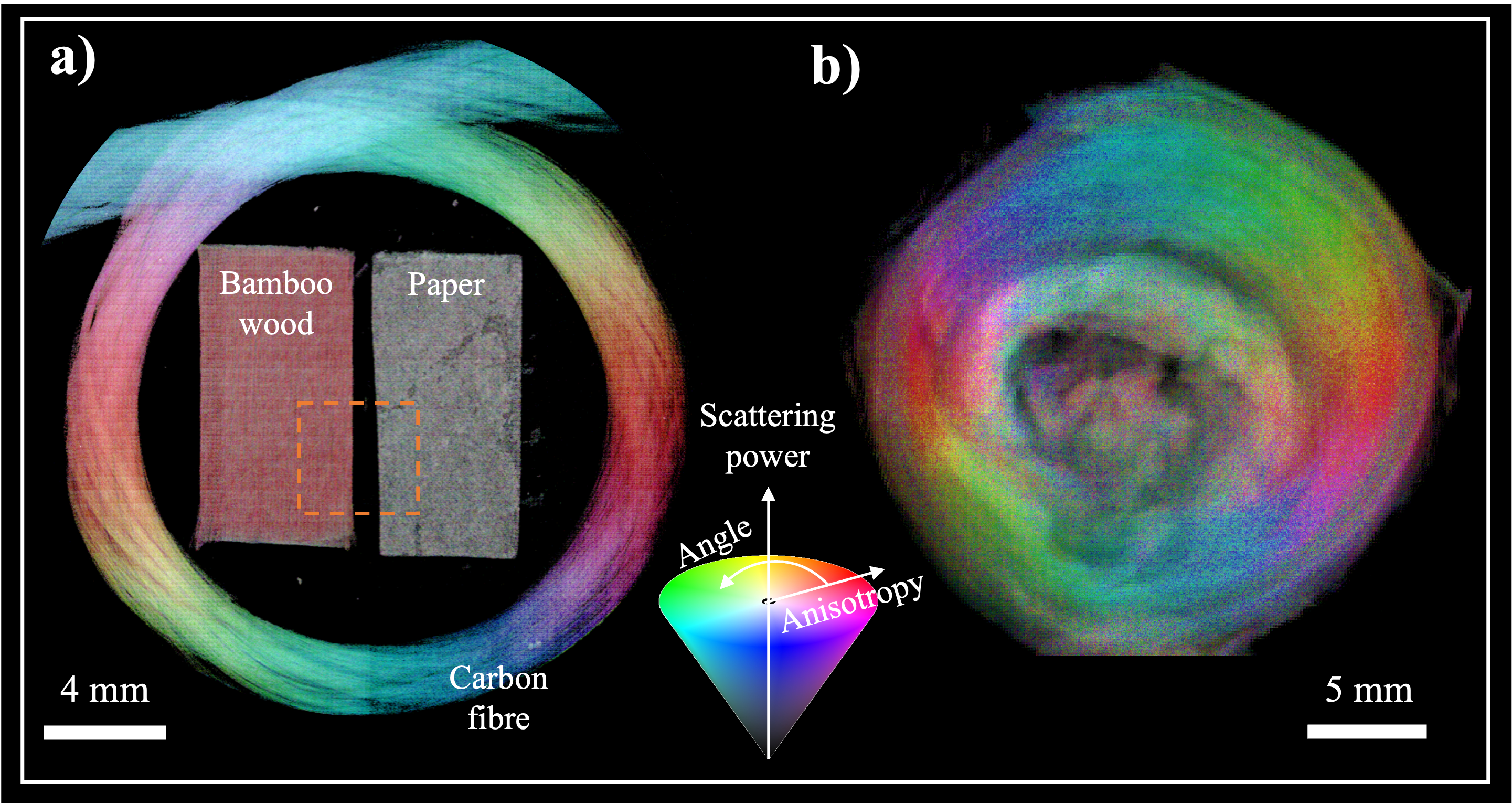}
\caption{\textbf{Directional x-ray dark-field radiography with the compact laboratory system.}
HSV colour mapping of directional scattering analysis showing primary scattering direction (Hue), anisotropy (Saturation), and scattering magnitude (Value) for (a) custom-built carbon fibre loop phantom, with wood and paper insets and (b) bovine intervertebral disc.}
\label{fig:directional_darkfield_2}
\end{figure*}

\section{Results and Discussion}

\subsection{Directional Dark-Field Imaging}

The system’s directional dark-field radiography capability is presented in Figure~\ref{fig:directional_darkfield_2}. The composite images of the custom-built phantom (a) and bovine intervertebral disc (b) use an HSV colour map, where hue encodes the primary scattering direction in the $(x,y)$ plane ($\theta$, Eq. \ref{eq:theta}), saturation represents the anisotropy coefficient ($\kappa$, Eq. \ref{eq:kappa}), and value corresponds to the scattering magnitude ($\lambda_{\text{max}}$, Eq. \ref{eq:kappa}). This representation allows to synthesise information about the magnitude, the direction as well as whether the scattering exhibits a preferential direction, in a single image. The individual parameter maps and grayscale scattering images that were used to generate these composite views are provided in the Supplementary Material.

The phantom, composed of a loop of carbon fibres, bamboo wood, and paper, demonstrates how directional scattering reflects underlying microstructural organisation (Figure~\ref{fig:directional_darkfield_2}a). In the carbon fibre bundle, which curves into a closed loop, the retrieved scattering direction follows the local fibre orientation, producing a smooth hue gradient, that repeats itself at opposite positions around the loop, where the fibres are aligned along the same direction. The bamboo wood shows predominantly horizontal scattering, which reflects the vertical orientation of its internal fibres. In contrast, the paper sample appears gray in the HSV representation. Whilst the presence of a structure finer than the system resolution is detected by a well-discernable signal, the analyis cannot attribute a specific colour. This indicates a microstructure composed of randomly oriented fibres, with no preferential direction observed.

Figure~\ref{fig:directional_darkfield_2}b shows the directional dark-field radiograph of a bovine intervertebral disc. The retrieved scattering directions follow the circumferential arrangement of the lamellae in the annulus fibrosus, consistent with the highly ordered collagen fibre architecture. In comparison, the central nucleus pulposus exhibits weaker anisotropy and less directional structure, reflecting its more homogeneous composition. The annulus fibrosus is made of concentric layers of mostly type I collagen fibres, typically in the range of $50-500$ nm that are organised in larger aggregates and bundles reaching few micrometers in diameter. The main function of these fibers and fibres structures is to provide tensile strength. In contrast, the nucleus pulposus is much less structured and the collagen is mostly of type II, characterised by thinner diameters $20-100$ nm and typically not organised in larger bundles. The primary function of this structure is to absorb compressive forces and allow for flexibility. Some of the main characteristics of these two distinct structures in the intervertebral disc are directly observable in the directional radiography, even if the structural differences are expressed at a length scale that is well beyond the spatial resolution of the system. The outer layer has a stronger signal and much better defined colours, the inner part exhibits a weaker signal, without a clear association between colours and direction. The ability of X-ray dark-field imaging to visualise this information on a large field of view enables the observation of the spatial distribution of these characteristics in a much wider context. These observations can be linked with the function of the underlying microstructure on a scale that is sufficiently large to represent the entire intervertebral disc. This ability to visualise local fibre orientation in unstained tissue may provide a valuable tool for studying early degenerative changes in intervertebral discs, which are often associated with microscopic disruptions that precede gross morphological alterations \cite{Zhao2007}.

\begin{figure*}[h]
\centering
\includegraphics[width=\linewidth]{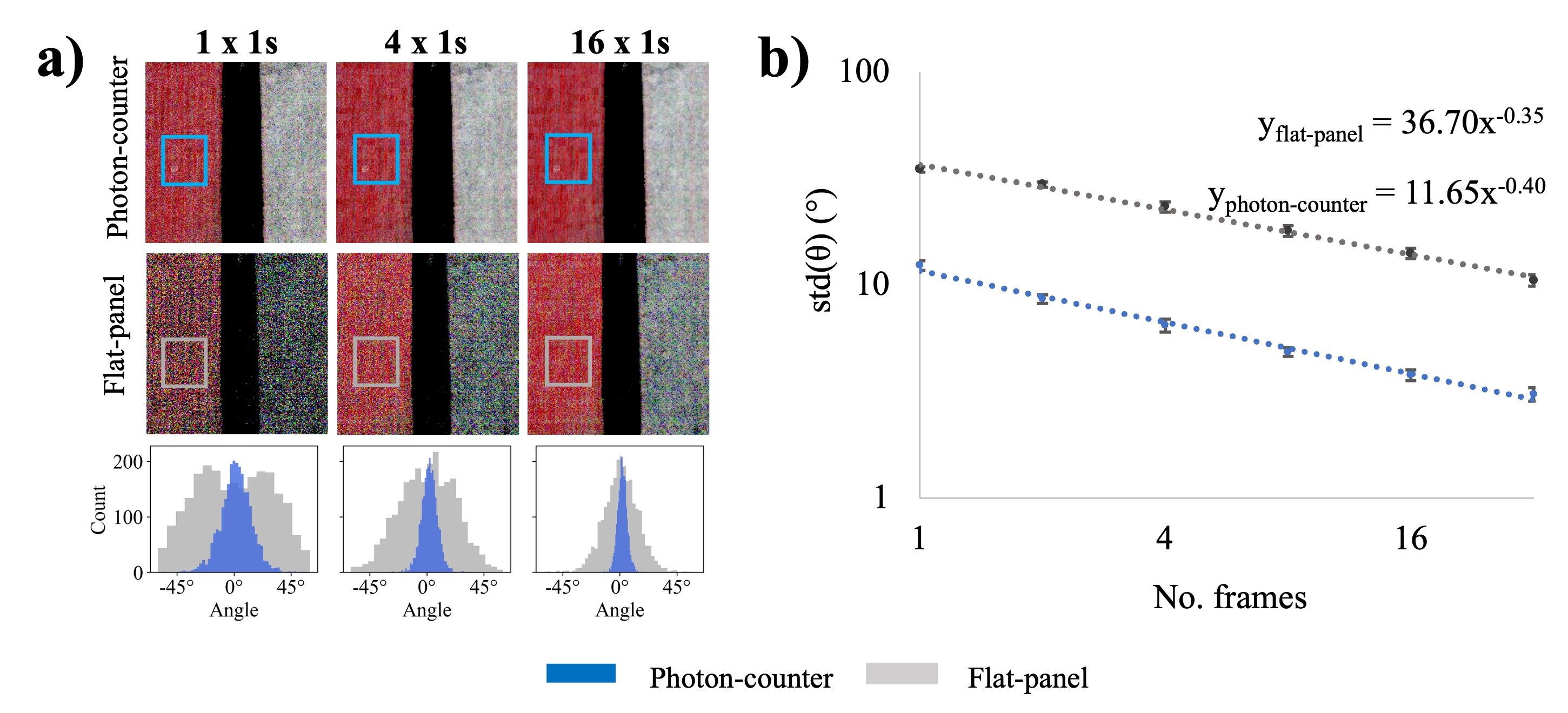}
\caption{\textbf{Directional dark-field signal characterisation.}
(a) Zoomed-in images of the ROI highlighted with the dashed square box of the phantom image in Figure \ref{fig:directional_darkfield_2}, showing directional dark field imaging of bamboo wood (left) and paper (right), at different exposure levels (columns) and for the two types of detectors we investigated (rows). The system's ability to retrieve the primary scattering orientation depends on the detector technology. This is qualitatively observed in the image noise and in the histograms of the hue component from the highlighted bamboo regions. The dispersion of the retrieved fibres' angles (hues in Figure \ref{fig:directional_darkfield_2}) appears substantially narrower when observed with a photon-counting detector (blue histogram) in comparison to a flat-panel (gray histogram). The trend of precision as a function of the exposure time for both detector technologies is shown in panel (b). A similar trend is observed in the two cases, however an improved performance is associated with the photon-counter.}
\label{fig:directional_darkfield_3}
\end{figure*}

The system’s ability to retrieve the dominant scattering orientation is influenced by the choice of detector technology, with the photon-counting detector showing higher precision. This is qualitatively visible in the reduced noise in the zoomed-in images and the hue ($\theta$ angle) histograms corresponding to the bamboo regions (Figure~\ref{fig:directional_darkfield_3}a). Here, the two detector technologies we investigated are compared to each other, and their precision is characterised as a function of the exposure time. Consistently higher precision is observed for the photon counter. To quantify angular precision, we evaluated the consistency of the retrieved scattering direction in a region of interest of $10 \times 50$ pixel region in the bamboo wood, highlighted with the blue and the gray squares in Figure \ref{fig:directional_darkfield_3}a. The angular precision, $\sigma(\theta)$, was defined as the standard deviation of the angles retrieved within the region. The analysis was repeated in five regions to obtain an average precision estimate. The highest observed angular precision was achieved with a 16 s exposure and the photon counter, and it was less than 1$^\circ$ (Figure \ref{fig:directional_darkfield_3}b). Longer exposures improved angular precision, with the dependence following a power law close to the ideal quantum-limited behaviour ($\propto t^{-1/2}$). Fits to the photon-counting and flat-panel data gave exponents of $-0.40 \pm 0.02$ and $-0.35 \pm 0.02$, respectively, indicating a mild deviation from the ideal $-0.5$. The observed deviation from the ideal behaviour was more pronounced for the flat panel. This is attributed to the dark-current accumulation across multiple frames, as well as sensor instabilities in the photon-counting detector. Additionally, residual variations in fibre orientation within the bamboo sample may contribute to the observed angular dispersion. These factors collectively limit the rate at which precision improves with exposure time. However, we observe that for exposure times in the region of $1-10$ seconds, the precision is largely dominated by the accumulated photon statistics. The effect of system geometry on angular precision is presented in the Supplementary Material (Figure 9), where we show that sensitivity improves with magnification but remains largely unaffected by total system length, hence supporting our choice for a compact implementation.

\begin{figure*}
\centering
\includegraphics[width=\linewidth]{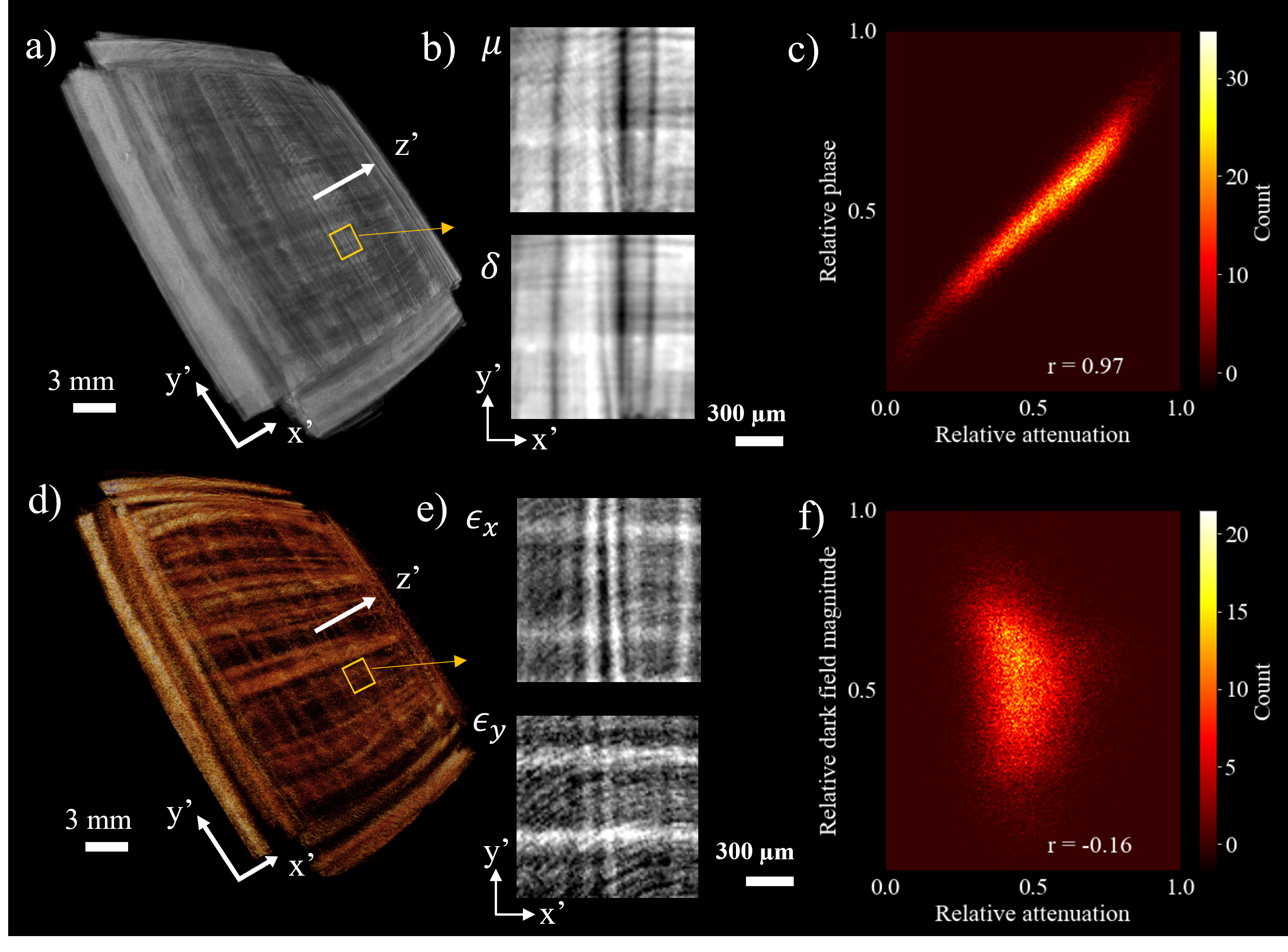}
\caption{\textbf{Multicontrast x-ray microtomography of fibre-reinforced composite.} 
(a,d) Volume renderings of the reconstructed linear attenuation coefficient $\mu$ and directional dark field coefficient $\epsilon_x$. 
(b,e) Zoom-in slices: attenuation ($\mu$) and phase ($\delta$) vary smoothly with laminate banding, whereas directional scattering ($\epsilon_x$, $\epsilon_y$) shows localised sub-voxel heterogeneity within the laminate. In this view, features approximately perpendicular to the system’s sensitivity direction appear stronger in the corresponding map: horizontal in $\epsilon_y$, vertical in $\epsilon_x$ (panel e). (c,f) Two-dimensional histograms from selected normalised sub-volumes show a strong voxel-wise correlation between $\mu$ and $\delta$, and a markedly weaker correlation between $\mu$ and the magnitude $\sqrt{\epsilon_x^2+\epsilon_y^2}$, indicating complementarity between the two channels.}
\label{fig:tomography_composite}
\end{figure*}

\subsection{Multi-contrast tomography}

Figure~\ref{fig:tomography_composite} shows tomographic reconstructions of attenuation, phase, and scattering signals from a fibre-reinforced polymer composite. Volume renderings of $\mu$ and the directional dark-field coefficient $\epsilon_x$ are presented alongside reconstructed slices of all four contrast channels aligned with the 0/90$^\circ$ fibre stacking sequence. The reconstructed volumes have an isotropic voxel size of $\sim$ $30 \times 30 \times 30$ ~\textmu m$^3$. In this view, $\mu$ and $\delta$ vary smoothly with visible laminate banding, whereas the scattering maps show localised, higher spatial frequency variations. These are consistent with sensitivity to sub-resolution heterogeneity within the laminate (e.g., fibre–matrix interfaces and bundle texture), although we did not independently verify the specific microstructural origin in this specimen. We further observe that regions where fibres appear perpendicular to the system’s sensitivity direction tend to show a stronger signal in the corresponding map: horizontal fibres appear brighter in $\epsilon_y$ and vertical fibres in $\epsilon_x$ (Figure~\ref{fig:tomography_composite}e), which is consistent with the expected angular dependence of dark-field contrast.

To assess the relationship between contrast mechanisms, two-dimensional histograms were computed from a $20 \times 40 \times 40$ voxel sub-volume located centrally within the sample. For comparison, voxel values in each channel were normalised to the [0, 1] range using min–max scaling within the subvolume. For quantifying the scattering, or dark-field signal, we computed the magnitude $\sqrt{\epsilon_x^2+\epsilon_y^2}$ to include the signals from both directions in the analysis. Attenuation and phase showed a strong voxel-wise correlation (correlation coefficient $R \approx 1$),  consistent with their shared sensitivity to electron density in what can largely be considered a constant $Z$ sample (Figure~\ref{fig:tomography_composite}c). In contrast, attenuation and dark field signals showed a markedly lower correlation ($R = -0.1$, Figure~\ref{fig:tomography_composite}f), illustrating the complementary nature of dark-field contrast in revealing structural features not visible in attenuation or phase images. We interpret the fundamental differences we obeserved between these two histograms  (Figure~\ref{fig:tomography_composite} panels c and f) as evidence that through X-ray dark-field tomography it is possible to gain new insights on the microstructure of the sample, that would not be accessible through the attenuation or the phase contrast channels. The main difference that can be observed between the attenuation and phase tomography (panel c) is a gain in signal-to-noise ratio: the same image features are represented with a smaller dispersion in phase than they are in attenuation. This could, at least in theory, be compensated through increased exposure and a collection of a higher X-ray statistic at the detector. The same considerations do not hold for panel f) where the correlation between the two contrast channels is much lower. In other words, given an attenuation or a phase tomography of a sample, it would be very difficult to predict how that sample appears under the dark-field contrast channel.

\begin{figure*}
\centering
\includegraphics[width=\linewidth]{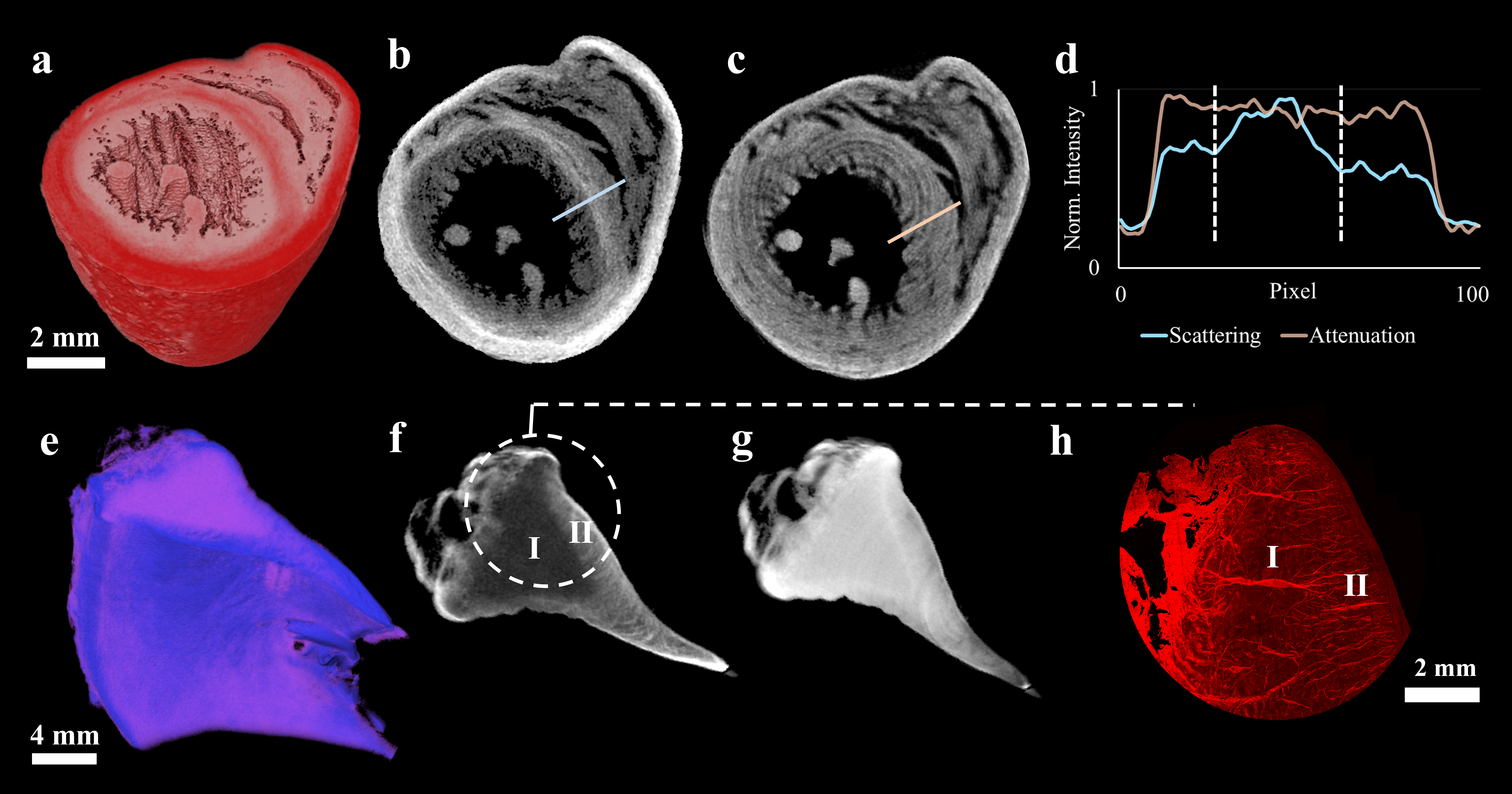}
\caption{\textbf{X-ray attenuation and dark-field microtomography of murine heart (a–d) and porcine meniscus (e–h).} 
(a,e) Volume renderings of the reconstructed dark-field magnitude $\sqrt{\epsilon_x^2 + \epsilon_y^2}$ with isotropic resolution show structural differences across tissue regions. 
(b,f) Axial slices of the dark-field channel show spatial variations in the reconstructed signal arising from microstructure differences within the tissue. 
(c,g) Corresponding axial reconstructions of the linear attenuation coefficient $\mu$ show comparatively uniform signal across the same regions. 
(d) Intensity profiles from the endocardium to the epicardium in the heart reveal a sub-surface layer visible primarily in the dark-field signal, linked to changes in fibre organization along this profile. 
We observe a transition in fibre density from the outer to the inner meniscus, which correlates with the contrast differences observed in the dark-field signal (f) between regions marked (I) and (II), not apparent in attenuation (g). (h) Multi-photon fluorescence microscopy of picrosirius red-stained tissue section of the meniscus after rehydration shows the collagen fibre organisation within the tissue.}
\label{fig:tomography_tissues}
\end{figure*}

We demonstrate the system’s capability for volumetric imaging of soft tissues using two ex vivo biological models: a murine heart and a porcine meniscus. Figure~\ref{fig:tomography_tissues} shows volume renderings of the dark-field magnitude $\sqrt{\epsilon_x^2 + \epsilon_y^2}$ (Figure~\ref{fig:tomography_tissues} panels a,e), alongside axial slices of the dark-field signal (Figure~\ref{fig:tomography_tissues} panels b,f) and the linear attenuation coefficient $\mu$ (Figure~\ref{fig:tomography_tissues} panels c,g). The dark-field magnitude combines the directional scattering components to improve the signal-to-noise ratio.

In the murine heart (Figure~\ref{fig:tomography_tissues}a), the scattering magnitude shows a bright myocardial layer (Figure~\ref{fig:tomography_tissues}b), which is not observed in the attenuation reconstruction (Figure~\ref{fig:tomography_tissues}c). This layer is also evident in the intensity profiles acquired from the endocardium to the epicardium, presented in Figure~\ref{fig:tomography_tissues}d. We attribute this feature to the mid-myocardial region, where a dense population of circumferentially aligned fibres generates the dark-field signal, consistent with observations from other imaging techniques \cite{GonzalezTendero2017, Gilbert2012}. Myocardial fibres in rats are approximately 10 \textmu m in diameter and beyond the resolving power of the attenuation and the phase tomography channels. Nonetheless, the dark-field tomography associates different intensities to different layers in the heart, enabling an identification of the boundaries between endocardium, myocardium and epicardium (Figure~\ref{fig:tomography_tissues}d).

The porcine meniscus scattering image (Figure~\ref{fig:tomography_tissues}e) shows a clear spatial variation across the tissue, with a stronger signal visible in the outer region (labelled II in Figure~\ref{fig:tomography_tissues}f). This contrast gradient is less apparent in the corresponding attenuation slices (Figure~\ref{fig:tomography_tissues}g). To investigate the underlying structure, we performed histological analysis using Picrosirius red staining after tissue rehydration and sectioning, as detailed in the Supplementary Material. Imaging with multi-photon excitation revealed a higher density of collagen fibres in the outer meniscus (region II) compared to the inner region (I), as shown in Figure \ref{fig:tomography_tissues}h. Picrosirius red binds selectively to collagen and exhibits strong fluorescence under multi-photon excitation, allowing high-resolution imaging of the architecture of collagen fibres. The increased scattering signal in region II is attributed to lamellar collagen tie fibres that run parallel to the meniscus surface at the femoral and tibial contact points. These fibres extend radially into the tissue, gradually merging with the circumferential collagen network that dominates the inner region \cite{Petersen1998, Andrews2014}. However, the inner meniscus exhibits weaker scattering, consistent with its more cartilaginous composition and reduced collagen fibre alignment. These results highlight the ability of dark-field imaging to map tissue microstructure, offering a contrast channel that is driven by fibre density, alignment and orientation within a voxel, rather than an overall quantification of the tissue attenuation or phase shift associated with that voxel. In this sense, dark-field micro-tomography offers a way to gain insights about the sample's micro-structure at scale lengths that are well below the spatial resolution or the voxel size. At the same time, these features can be observed in a larger field of view, providing an extended context to these observations.

\section{Conclusions}

We have presented a compact, laboratory-based x-ray imaging system that enables the retrieval of attenuation, phase, and directional dark-field signals using two-directional beam tracking (2DBT). The system requires minimal optical components and accommodates standard x-ray sources and detectors in less than 1~m total distance, offering a practical route to dark-field imaging and directional scattering analysis in settings where source coherence and instrumentation footprint are limited. We demonstrated its capabilities on a range of samples, revealing features of their fibrous and heterogeneous microstructure. Dark-field imaging and directional scattering analysis provide complementary structural information that generates at scale lengths that are beyond the imaging system's spatial resolution, and which are not captured through conventional attenuation or phase tomography.

Our results highlight the potential of this approach for the non-destructive characterisation of fibrous media, including engineered composites and notably biological soft tissues. We believe that these capabilities can enable new applications in non-destructive testing, tissue characterisation, pathology, and other applications in biology and medicine.

\section*{Acknowledgments}
This work was supported by the Wellcome Trust (221367/Z/20/Z); the Francis Crick Institute, which receives its core funding from Cancer Research UK (CC0103), the UK Medical Research Council (CC0103), and the Wellcome Trust (CC0103); and by the Engineering and Physical Sciences Research Council (EP/T005408/1, EP/T02593X/1, EP/S021930/1), through the Nikon–UCL Prosperity Partnership on Next-Generation X-ray Imaging, the National Research Facility for Lab X-ray CT (NXCT), and the UCL Centre for Doctoral Training in Intelligent, Integrated Imaging in Healthcare (i4health).
CH is supported by the Royal Academy of Engineering under the Research Fellowship scheme.
AO is supported by the Royal Academy of Engineering under the Chair in Emerging Technologies scheme (CiET1819/2/78).
We are grateful to Prof. Paul Fromme for insightful discussions on the tomographic imaging of the composite sample.
We thank the Crick Advanced Light Microscopy Science Technology Platform for assistance with microscopy imaging, and the Crick Experimental Histopathology Science Technology Platform for their support with sample preparation, staining, and histological analysis of the meniscus tissue.

\section*{Author contributions}
ME and CN-L conceived and designed the study. CN-L, AJG-G, HA, AD, AP, RD’A, AA, and SC carried out the experiments. CN-L, AJG-G, AA, SC, AD, HA, and ME developed the computer code and performed data analysis. AO, DB, SC, CHK, and ME secured funding for this work and provided access to facilities. ME, DB, SC, CHK, and AO supervised the analysis and data interpretation. CN-L and ME wrote the manuscript with contributions and critical revisions from all authors. All authors reviewed and approved the final version of the manuscript.

\section*{Data availability}
The datasets generated and analysed during the current study are available from the corresponding author upon reasonable request.

\section*{Code availability}
The custom code used for beam-tracking signal retrieval, tomographic reconstruction, and data analysis is available from the corresponding author upon reasonable request.

% \bibliographystyle{unsrtnat}
% \bibliography{sample}

\clearpage

\appendix
\section*{Supplementary material}

\section{Directional dark-field data analysis}

This section complements Section 2.1 of the main text by illustrating the directional dark-field retrieval process with experimental images and intermediate signal maps. The intermediate maps include the variance and covariance images, as well as the derived orientation, anisotropy, and scattering magnitude maps that are subsequently combined into the HSV rendering shown in the main manuscript.

\begin{figure*}[b!]
\centering
\includegraphics[width=\linewidth]{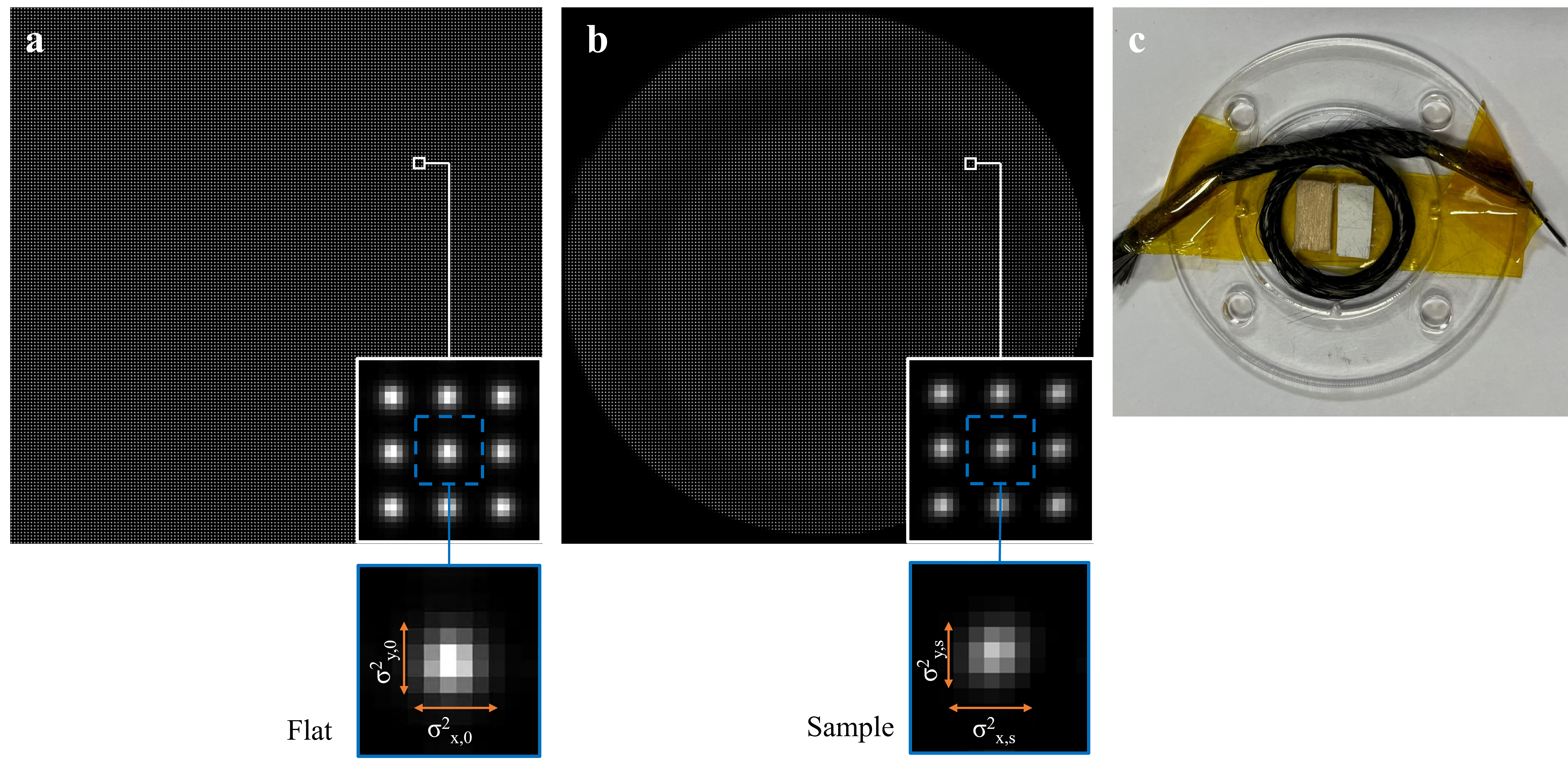}
\caption{\textbf{Dark-field signal retrieval.} Experimentally acquired images without (a) and with (b) the sample in place. (c) The sample is a custom-built phantom composed of bamboo, carbon fibres, and paper. Insets show zoom-ins into a $3\times3$ region of beamlets located within the carbon fibre area. Although the effect of the sample is barely visible by eye, scattering is quantified by comparing spatial variance between flat-field and sample-exposed beamlets.}
\label{fig:experiment}
\end{figure*}

Figure \ref{fig:experiment} shows raw experimental images of the custom-built phantom, with and without the sample in place. While the effect of scattering is barely visible by eye, it is quantified by comparing beamlet variances and covariances as defined in Section 2.1 of the main text.

\begin{figure*}
\centering
\includegraphics[width=0.8\linewidth]{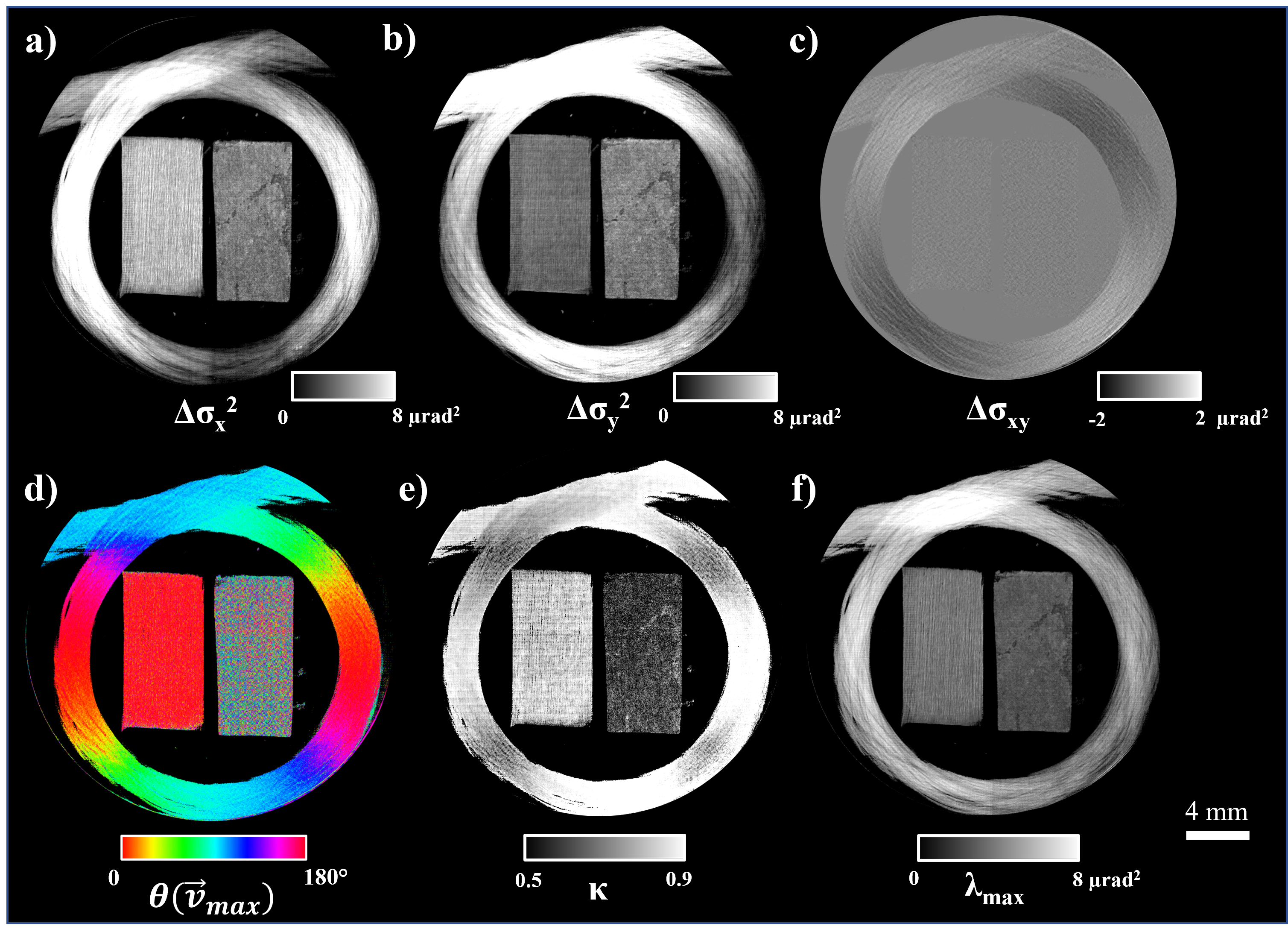}
\caption{\textbf{Directional dark-field imaging of custom-built phantom.} (a–c) Maps of the scattering variances in the $x$ and $y$ directions ($\Delta\sigma_x^2$, $\Delta\sigma_y^2$), and their covariance ($\Delta\sigma_{xy}$). (d–f) Directional parameters derived from the covariance matrix: principal scattering orientation $\theta$, anisotropy coefficient $\kappa$, and maximum scattering magnitude $\lambda_{\text{max}}$. These intermediate maps illustrate how the final HSV rendering (main text, Fig. 2) is synthesised.}
\label{fig:phantom_analysis}
\end{figure*}

The resulting maps are shown in Figure \ref{fig:phantom_analysis}. Variances along $x$ and $y$ and the covariance ($\Delta\sigma_x^2$, $\Delta\sigma_y^2$, $\Delta\sigma_{xy}$) form the basis of the covariance matrix (see Eq. 3 in the main text). From these, the principal scattering orientation $\theta$, anisotropy coefficient $\kappa$, and maximum scattering magnitude $\lambda_{\text{max}}$ are obtained by eigen-decomposition.

\begin{figure*}
\centering
\includegraphics[width=0.8\linewidth]{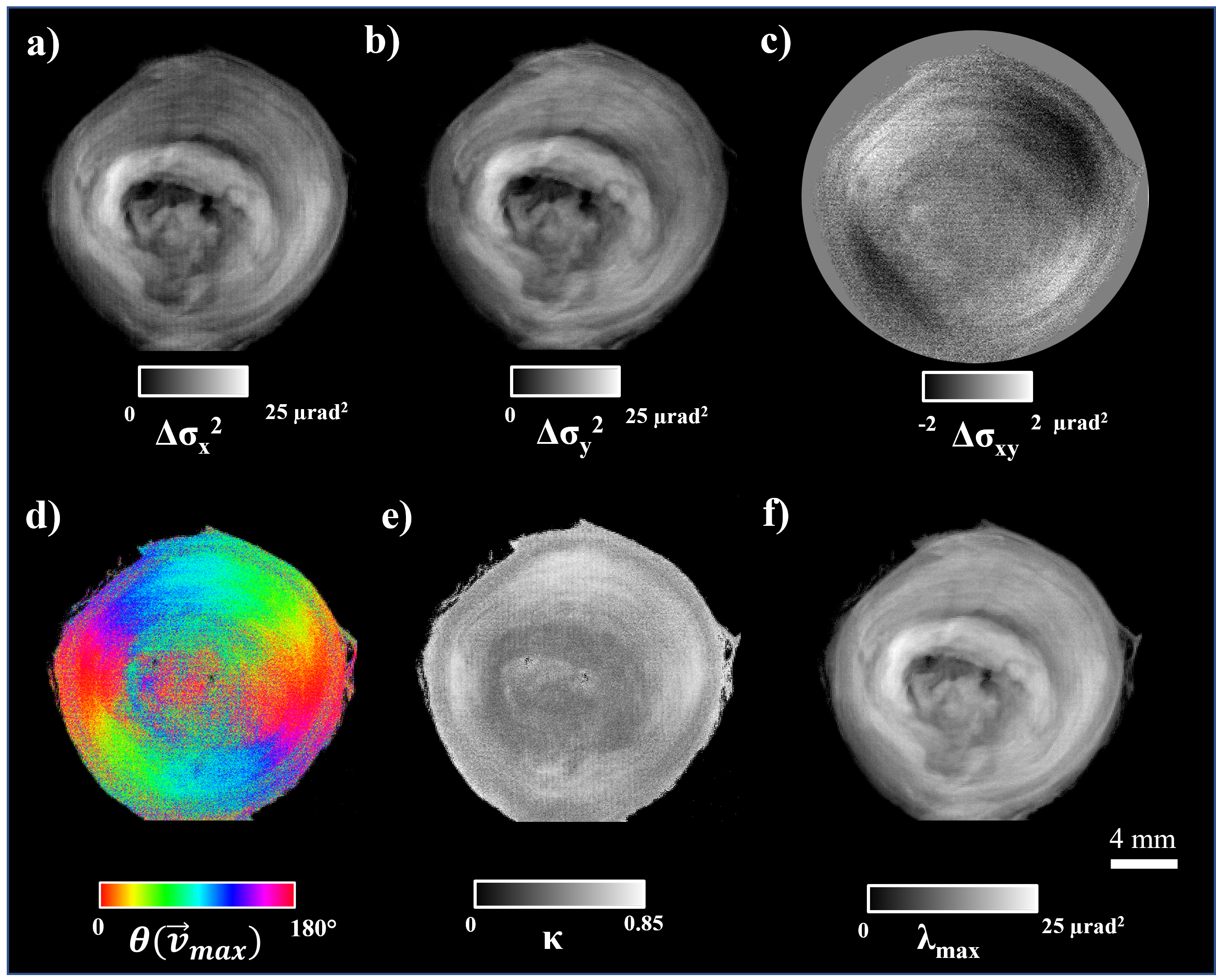}
\caption{\textbf{Directional dark-field imaging of bovine intervertebral disc.} (a–c) Maps of the scattering variances in the $x$ and $y$ directions ($\Delta\sigma_x^2$, $\Delta\sigma_y^2$), and their covariance ($\Delta\sigma_{xy}$). (d–f) Directional parameters derived from the covariance matrix: principal scattering orientation $\theta$, anisotropy coefficient $\kappa$, and maximum scattering magnitude $\lambda_{\text{max}}$. These intermediate maps illustrate how the final HSV rendering (main text, Fig. 2) is synthesised.}
\label{fig:disc_analysis}
\end{figure*}

The scattering variance and covariance maps (Figures~\ref{fig:phantom_analysis}a–c) show how different structures within the phantom give rise to distinct directional scattering signals. For instance, stronger $\Delta \sigma_x^2$ is observed in vertically aligned bamboo, while diagonally oriented carbon fibres show a more pronounced covariance. The extracted directional parameters (Figures~\ref{fig:phantom_analysis}d–f) characterise local variations in the sample’s microstructure and scattering anisotropy. The orientation map reflects the local fibre alignment, as seen in the hue variation along the carbon fibre loop. The anisotropy map shows that bamboo and carbon fibres have high values of $\kappa$, in contrast with the paper sample, which shows low anisotropy despite comparable scattering magnitude. This difference is attributed to the more isotropic microstructure of the paper, which does not exhibit a dominant scattering orientation.

For comparison, we also extracted the same intermediate maps for a bovine intervertebral disc (Figure \ref{fig:disc_analysis}). The variance and covariance maps (panels a–c) show directional signals linked to their sensitivity axes: scattering is stronger in regions where fibres are aligned perpendicular to the $x$ or $y$ sensitivity direction, while the covariance highlights regions with diagonal orientation where the two components vary together. From these, the derived maps of orientation, anisotropy, and scattering magnitude (panels d–f) more clearly reveal the organised lamellar collagen architecture. The orientation map shows the expected circumferential alignment of fibres in the annulus fibrosus, while the anisotropy coefficient demonstrates a radial variation, with strong alignment in the outer lamellae and a progressive loss of anisotropy toward the centre. This transition reflects the two main tissue regions: the outer annulus fibrosus with its concentric lamellae, and the central nucleus pulposus, which exhibits weaker and more isotropic scattering consistent with its more homogeneous composition.

\begin{figure*}
\centering
\includegraphics[width=\linewidth]{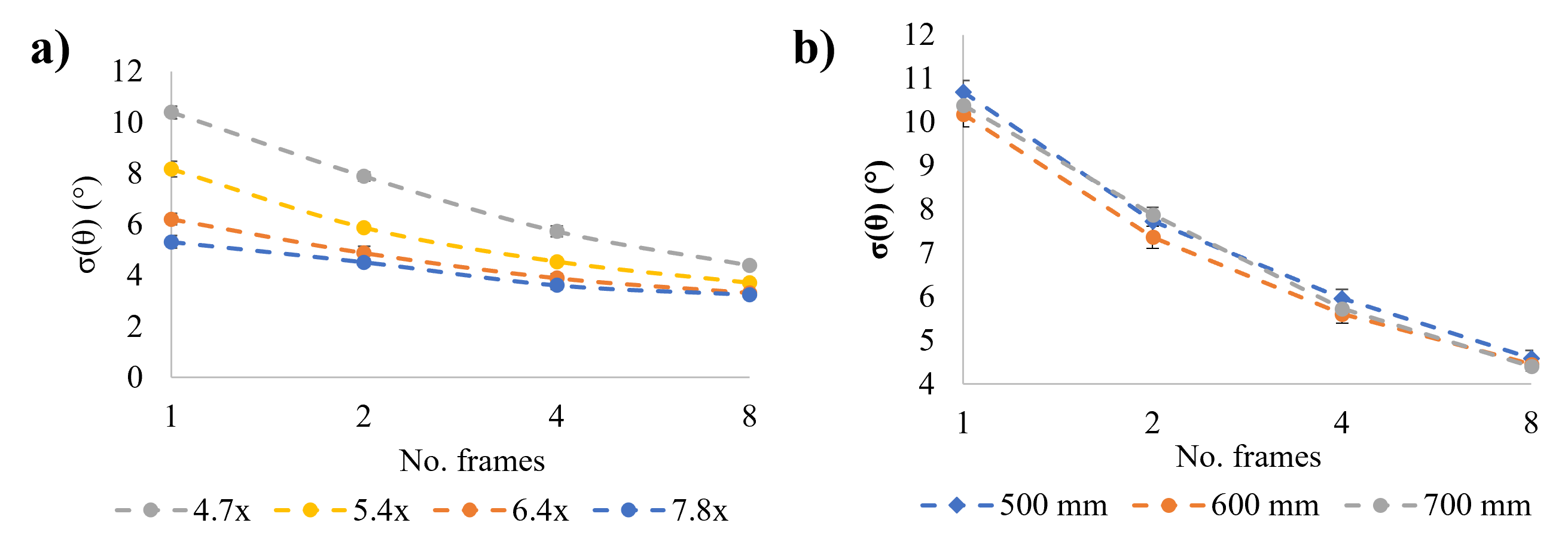}
\caption{\textbf{Directional dark-field signal angular precision.} (a) Angular precision ($\sigma(\theta)$) as a function of exposure time for three different magnifications at fixed source-to-detector distance (700~mm). Higher magnification improves sensitivity to beamlet deformation. (b) Angular precision versus exposure time for different system lengths, at constant magnification ($\sim$4.7$\times$), showing minimal dependence on total system length.}
\label{fig:signal_precision}
\end{figure*}

\section{Dark-field signal characterization}

The main text summarises how angular precision depends on detector technology and exposure time. Here, we further explore its dependence on system geometry.

Angular precision was defined as the standard deviation of the retrieved principal scattering direction ($\theta$) within a region of high fibre alignment in the scattering phantom. Measurements were performed in a $10 \times 50$ pixel region of interest located in the bamboo sample. The same region was used across all configurations for direct comparison.

Figure~\ref{fig:signal_precision}a presents angular precision for three different magnifications (3.3$\times$, 4.7$\times$, and 6.1$\times$), with the source-to-detector distance fixed at 700~mm. Higher magnification shows improved angular precision, which we attribute to the increased projected probe size. This enhances the sensitivity to the beamlet changes in shape and improves photon statistics per beamlet.

Figure~\ref{fig:signal_precision}b shows angular precision as a function of source-to-detector distance (550~mm, 700~mm, and 850~mm) at a fixed magnification of approximately 4.7$\times$. The retrieved orientation precision remains relatively stable across system lengths, indicating that magnification and exposure time are the dominant factors in determining measurement performance. The trade-off between improved photon statistics at shorter distances and reduced measurable broadening of the scattering signal due to geometry-dependent projection ($\propto 1/z_{od}$) explains the limited dependence of angular precision on total system length.

\section{Histology and light microscopy of porcine meniscus}

Following X-ray imaging, the HMDS-dried meniscus was immersed in 70\% ethanol for one week, then rehydrated over four weeks in a 30:70 solution of glycerol and industrial methylated spirits (IMS). In this solution, glycerol functioned as a cellular structure-preserving agent via hydrogen bonding interactions~\cite{MacLeod2018,Toba2023}, while IMS facilitated tissue rehydration in preparation for paraffin wax embedding, microtome sectioning, and subsequent histological staining~\cite{Lewis2016}.

Radial sections of 3 to 5 $\mu$m thickness were obtained using a ThermoFisher rotary microtome, then dewaxed and rehydrated using the Sakura Tissue-Tek® Prisma system. The rehydrated sections were incubated in Weigert’s haematoxylin solution for 10 minutes, washed in warm water, and subsequently immersed in Picrosirius red solution for one hour. Differentiation was performed using 0.5\% acetic acid, followed by final rinses in 100\% IMS and xylene for tissue clearing. Finally, the stained sections were mounted using Sakura Tissue-Tek® Glas™ mounting medium.
The tissue sections were imaged on a Zeiss LSM 880 NLO multiphoton microscope, illuminating the sample with a Coherent Chameleon near-infrared pulsed laser tuned at 800 nm, using a Plan Apochromat 20x/0.8 NA objective. The fluorescence signal was acquired with Non-Descanned Detectors equipped with a bandpass filter for picrosirius red (BP 575-610 nm) and a shortpass filter for Second Harmonic Generation (SP 485 nm). 
% Bibliography

% \bibliographystyle{unsrtnat}
% \bibliography{sup_sample}

\end{document}